\documentclass[12pt]{amsart}
\usepackage{amsmath,amssymb,amscd,enumerate,bbm}
\newcommand{\titl}{SPECTRA AND STRAINS}
\title{{\titl}}
\author{V. Golyshev}

\newcommand{\cal}{\mathcal}

\def\R{{\Bbb{R}}}
\def\C{{\Bbb{C}}}
\def\Q{{\Bbb{Q}}}
\def\Z{{\Bbb{Z}}}

\def\E{{\cal{E}}}

\newcounter{pphcounter}[section]
\renewcommand{\thepphcounter}{\thesection.\arabic{pphcounter}}
\newcommand{\pph}{\bigskip \refstepcounter{pphcounter}
    \bf  \thepphcounter. \rm\quad}

\makeatletter
\@addtoreset{equation}{pphcounter}
\makeatother

\newcommand{\dimens}{{\mathrm{dim \;} }}

\def\A1{{{\Bbb{A}}^1}}
\def\P1{{{\Bbb{P}}^1}}

\newcommand{\Qlbar}{{\overline{\Q} _l}}

\newcommand{\Spec}{{\text{Spec }}}
\def\Anr1{{{\Bbb{A}} (n,r+1)}}

\def\Gm{{\bf G_m}}

\def\SL2{{\mathrm SL2}}

\renewcommand{\epsilon}{\varepsilon}

\renewcommand{\P}{{\Bbb{P}}}
\renewcommand{\phi}{{\varphi}}

\newcommand{\lto}{\longrightarrow}

\newcommand{\inv}{{\mathrm{inv}}}

\newcommand{\spec}{\mathrm{spec}\,}
\newcommand{\Gal}{\mathrm{Gal}\,}

\date{\today}

\begin{document}

\begin{center}
\maketitle

\end{center}

\bigskip

\parbox{340pt}{\small \bf Abstract. \rm This is a
blend of two informal reports on the activities
of the seminar on Galois representations and mirror symmetry given at the Conference
on classification problems and mirror duality at the Steklov Institute,
in March 2006, and at the Seminar on Algebra, Geometry and Physics
at MPI, in November 2007.
We assess where we are on the issue of the spectra of Fano varieties, and state problems.
We
introduce higher dimensional irreducible analogues of dessins,
the low ramified sheaves,
and hypothesize that Fano spectra relate to their geometric conductors. We
give a  recipe to a physicist.}

%
\bigskip
\bigskip
\bigskip

\section{The Fano Spectra}

\pph{\bf Spectra and anticanonical spectra.} Let $F$ be a Fano variety of index $d$, so that
$-K=dH$.
Consider the matrix of quantum multiplication by $H$. It has entries
in $\Q[q_i,q_i^{-1}]$, where $i$'s correspond, as usual, to
the numerical classes of curves on $F$.

One may specialize the matrix to $M_H$ in $\mathrm{Mat }\,(\Q [t,t^{-1}])$
by
$$q_i \mapsto t^{\text{degree of curve of class } i},$$
where the degree is
taken with respect to $H$. There is no need to do that when
$H^2(F)=\Z$ which we will freely assume henceforth. Let
$\inv$ stand for the multiplicative inversion $\Gm \lto \Gm$

We  define [provisionally!] the \emph{spectrum} to be the inv of the closed subscheme of
$\Gm = \Spec \Q [t,t^{-1}]$ given by the principal ideal generated by
$\det (M_H)$.

A concurrent notion is that of an \emph{anticanonical spectrum.} This arises in
a similar way when
one specializes the matrix of quantum multiplication by $-K$
to $M_{-K}$ in $\mathrm{Mat }\,(\Q [t,t^{-1}])$
by
$$q_i \mapsto t^{\text{degree of curve of class } i},$$
where the degree is now
taken with respect to $-K$. Up to a shift on the torus, the anticanonical spectrum is
the pullback of the spectrum under the $d$-isogeny
$\Gm \lto \Gm.$ One also considers the \emph{complete anticanonical spectrum} that
comprises all singularities of the regularized anticanonical quantum
$D$-module \cite{Go} by adding a suitable component at infinity.

\bigskip

\pph{\bf Quantum Lefschetz and stability of spectra.}
Givental's Quantum Weak Lefschetz theorem
implies that the spectra of Fanos are stable under
hyperplane sections: if $V$ is a hyperplane section of $F$ of index $>1$, then
$\spec F$ coincides with $\spec V$ up to a multiplicative shift. If the index of $V$ is $1$,
there exists a so--called Givental constant $g$ such that
$\spec F - g$ coincides with $\spec V$ up to a multiplicative shift, and from now on
we adjust our
definition of $\spec F$ to be what formerly was $\spec F -g$.


\pph{\bf Strains.} Two [deformation classes of]
Fano varieties are said to be in the same \emph{strain} if one is a hyperplane
section of the other. We extend that to an equvalence relation and define the spectrum
of a strain to be the spectrum of
any of its members. The spectrum is well defined up to a multiplicative shift.

\bigskip

If $V$ is a hyperplane section of $F$, we sometimes refer to $F$ as an \emph{unsection}
of $V$.

\pph{\bf Progenitors.} We call a Fano variety a \emph{progenitor} of its strain if
any variety in the strain
is its hyperplane section. In particular, progenitor have no unsections.

\pph{\bf Problems.} Given a Fano $F$, determine whether it is a
progenitor of its strain. Given a strain $S$, determine whether it is finite.

{\bf Example.} The strain of complete intersections in projective
space is infinite and has no progenitor.

\bigskip

One may choose to work with ``easier'' Fano varieties, or
their strains:

\pph{\bf Cellular, minimal, Tate.} A Fano variety $F$ is said to be \emph{minimal}
 if its cohomology
is as small as it can be ($H^{2k+1}(F)=0,H^{2k}(F)=\Z$). A Fano is said to be \emph{Tate}
if its motive has no non-Tate constituents. A Fano $F$ is said to be \emph{cellular} if
$F$ is a union of affine spaces:
$F=\bigcup {\mathbb{A}}_j^{i(j)}, \;{\mathbb{A}}_{j_1} \bigcap{\mathbb{A}}_{j_2}
= \varnothing \text{ if } j_1 \ne j_2.$  A strain is
called cellular/minimal/Tate if it has a cellular/minimal/Tate
variety in it.

\bigskip

\bf Examples. \rm The strain of complete intersections in projective
spaces is cellular and minimal and Tate. Its spectrum is one point defined
over $\Q$. A less trivial example is the strain of the
Grassmannian $G(2,5)$. It is again
cellular and minimal (its triple hyperplane section is a minimal Fano threefold
$V_5$) and Tate. Its spectrum consists of the two roots of
$t^2-11t-1$. Hence, $V_5$ is not a complete
intersection in a projective space. Spectra of rank $1$ Fano threefolds $F$ are
formed by elliptic points and cusp points on $X_0(N)/W_N$ where
$\displaystyle{N=\frac{(-K_F)^3}{2d^2}}$ \cite{Go}.

\pph{\bf Problem.} Clearly, cellular implies Tate, and minimal implies Tate.
Are there non-cellular minimal strains?

\pph{\bf Problems.} Find the spectra of all Fano / all cellular / all
minimal strains. Find the Fano spectral field  ${\mathcal F} \subset \overline{\Q}$,
is the minimal field of definition of all Fano spectra.

\pph{\bf Theorem.} \cite{GG} The spectra of all Grassmannians  are
defined over $\Q ^{ab}$.

\bigskip

Contrary to some expectations,
S. Galkin and I have found that it is possible, even in a case of a classical group,
to find in the spectrum of a generalized grassmannian an irreducible component (i.e.
a $\Gal \Q$ orbit)
whose Galois group is the symmetric group $S_n, n \ge 5$. We have also found
that two generalized grassmannians of different classical groups
may have a common non-trivial irreducible component in their spectra.
This shows that the spectrum
is a fine invariant and raises the following

\pph{\bf Problems.} To what extent does a spectrum determine its
(cellular) strain? Is it possible for two different (say, cellular) strains
to have the same spectrum?  And, vaguely: let $P_1$ and $P_2$ be the progenitors
of the strains $S_1$ and $S_2$. Let the intersection of $\Spec S_1$
and $\Spec S_2$ be non-empty and non-trivial. Does it imply that
there is a natural correspondence between $P_1$ and $P_2$?

\bigskip

\section{Low ramification sheaves and LRS spectra}

Is the way the cells of a
cellular variety are joined together  controlled by a dessin--type combinatorics?
Dessins should be generalized to positive relative dimensions as
``maximally Szpiro'' objects, such as flat morphisms
that have as few critical points as possible. We discuss this naive approach
in greater detail in
the next section;
meanwhile, to fix ideas, we
deal with [absolutely geometrically] irreducible Galois representations
of the field $\Q(t)$ that are low--ramified geometrically.

\pph{\bf Geometric ramification.} Let $U \subset \P ^1 / \C$ be a Zariski open subset, $S=\P ^1 -U.$ Let $L$
be a rank $r$  non-trivial irreducible  polarized local system over $U,$
i.e. a representation
$\phi : \pi_1(U^{an}) \lto O(r)/Sp(r),$ and let $L_x$ be its generic
fiber.
Its {\it ramification} is
$$R(L) = \sum_{s \in S(\C)} \dimens L_x / L_x^{I_s}.$$

\pph{\bf Low ramified local systems and their conductors.} A local system $L$
as above  is said to be
low ramified if $$R(L) = 2\, \text{rk}\, L.$$
Its geometric conductor is
the respective closed subscheme of  $\P ^1 \mid_{\C}$: the union of
points ${s \in S(\C)}$ each taken with multiplicity $ \dimens L_x / L_x^{I_s}.$

We want to consider the cases when $L$ is ``of arithmetic nature''. One may imagine
the following simplistic picture. First, one wants
the image of the monodromy to be in $GL(r,\Z)$.
Given a prime number $l$, one arrives at a tower of unramified Galois covers $U^{(l^n)}$
of
$U$ by considering the respective  system of $\mod l^n$ representations.
Assume $S$ is defined
over $\Q$, that is, $U$ is defined
over $\Q$. The second requirement is that the covers $U^{(l^n)}$ be defined over
$\Q$.

One may budge a low ramified local system on the complex analytic base
isomonodromically by shifting points in $S$, but  that will result, in general, in
a loss of definability of
the system of level covers over $\Q$, or in fact over any fixed number field (although  each
$U^{(l^n)}$ is definable individually over some number field by Weil).

A typical example of an ``arithmetic'' local system is an irreducible constituent
in the local system of relative cohomologies in a smooth pencil over $U\mid_{\Q}$.
As is usual in this context, we allow for monodromies in the ring of algebraic integers,
and for the finite base field change $K/\Q$:

\pph{\bf Low ramified $l$-adic sheaves.} A [necessarily tame]
lisse [absolutely] geometrically irreducible
$\Qlbar$-sheaf $\mathcal{L}$ on $U_{K}$ is said to be \emph{low ramified},
or to be \emph{an LRS},
if its geometric ramification computed as above
satisfies $R ({\mathcal L})= 2\, \text{rk}\, {\mathcal L}.$ We say that
an $l$-adic sheaf on $\P^1 \mid _K$ is low ramified if its restriction to its
ouvert de lissit\'e is low ramified. We define the \emph{conductors
of low ramified sheaves} as above.

Conductors of low ramified sheaves will also be referred to as \emph{LRS spectra}.

\pph{\bf Problem.} Find the conductors of low ramified
sheaves on $\P^1 \mid_\Q$ (resp. $\P^1 \mid_K$) of a given rank.

\pph{\bf Fano spectra and LRS spectra}. One may ask what the two worlds
have in common. Our interest in the LRS spectra arose from the fact
that the Landau-Ginsburg models of the rank $1$ threefolds are, motivically,
twisted Kuga--Sato families. \footnote{LG's of [quantum minimal] del Pezzos have modular
meaning, too.} The due generalization of such a Kuga-Sato or a
modular elliptic surface
over a rational base to a higher relative dimension $N-1$ is a pencil
$\pi : \E \lto \P^1 \mid_\Q$ such that the ``essential'' constituent of
$R^{N-1} \pi_!(\Qlbar)$ is a low ramified sheaf on the base. Consider now a minimal
Fano $F$; its completed anticanonical spectrum
is given by the symbol of the ``counting
equation DN'' of $F$. Now, a generic equation DN has been shown \cite{GS}
to be of low ramification,
in the sense that the local system of its solutions is. A conjecture of mirror symmetry
asserts that the counting DNs are of Picard-Fuchs type, hence, modulo the conjecture,
the spectrum of a ``generic'' minimal Fano is also an LRS spectrum.

\pph{\bf Problems.} On some genericity assumption on the variety
\footnote{A suitable quantum analogue of
absence of primitive algebraic classes in the middle dimension.},
is every component (= Galois orbit) of the
completed anticanonical spectrum of
a Fano/cellular Fano/Tate Fano also
a component of some LRS spectrum?  The reverse, `is every component of an LRS
spectrum also a part of some cellular spectrum?', is most probably refutable
as stated, but might become a real one if the premise is made a bit more specific
(prescribing types of some of local monodromies, etc).

\pph{\bf Arithmetic conductors of geometric conductors.}
We finish this section with the following observation:
the fields of definition of the components of the Fano spectra
tend to have small discriminants per degree.
The Galois group of the Fano threefold $V_{22}$
is $S_3$. The discriminant is $-44$.
The spectrum of the blowup of $\P ^3$ along $\P^1$
consists of two irreducible pieces. The Galois group of each is
$S_3$. The fields of definition are unramified over the
respective quadratic extensions. The discriminants are $-23,-31$,
exactly the two lowest
levels at which there emerge weight $1$ cuspforms.
Can the assertion that `the components
of the spectrum of a cellular variety are not too ramified' be made precise?
Are they close to the border allowed by the explicit formulae \footnote{Diaz y Diaz.}?
Rephrasing, shall we expect the combinatorics of the affine cells
to possess certain optimality properties? And on the LRS side, can one show
that the arithmetic conductors of geometric conductors of low ramified sheaves
are small? Vague as it is, this observation, if extended, may have
very practical consequences for the search for the spectra.

\bigskip

\section{The search}

In order to tabulate low ramified sheaves of low ranks,
one may proceed by fieldworking for another closely related zoology.

\pph{\bf The special Laurent polynomials zoo.} Let $M$ be the standard
lattice in $\R ^{N}$, and let $P$ be a Calabi-Yau lattice polytope, that is,
one with $1$ strictly internal point. Inasmuch as it is
allowed by $P$, a generic non--zero polynomial $\pi$ `tends' to be Morse(--Lefschetz), i.e.
have simple singularities and critical values. What we need is the opposite of the
generic: stratify $X$ according to
how the critical values come together and single out the Artinian strata. We refer
to the geometric
points in these strata as \emph{the special Laurent polynomials}~\footnote{The definition
of the stratification on $X$ seems to require a
good deal of local to global algebra
that can account for multiple singularities, non--isolated singularities and
singularities at the infinity of the compactification. It is not improbable, though,
that these fine contributions may add up to a practicable total.}.

\pph{\bf LRS vs Special Laurent.} One should not expect
the two classifications to directly translate
one into the other.
Which low ramified sheaves arise then as irreducible constituents in $R^{N-1} \pi_! (\Qlbar)$
with $\pi$ special Laurent? What are the conditions on $P$ that guarantee that
the special Laurent polynomials with support in $P$ produce low ramified sheaves,
or at least
some of them do?

According to Batyrev's idea
of mirror symmetry for non-torics,  a Fano $F$, say of Picard rank $1$, may degenerate
to a toric whose fan's unit vectors are the vertices of a polytope of CY type.
We may look for the weak Landau-Ginsburg model \cite{Pr} of $F$
in the linear space $X$ of Laurent polynomials with support in $P$. One should not
therefore be too
surprised \footnote{Upcoming
is S.~Galkin's thesis where some of these matters are worked out for $N=2,3$.}
to find inherent structural similarities between the [subdivisions within]
classifications of Batyrev
type degenerations, the low ramified sheaves and the special Laurent polynomials.

\end{document}